\documentclass[aps,showpacs,preprintnumbers,amsmath,epsf,amssymb,epsfig,nofootinbib]{revtex4}
\usepackage{bm}
\begin{document}

\title{Gravitational waves during inflation in presence of
a decaying cosmological parameter from a 5D vacuum theory of gravity}
\author{$^{1}$Silvina Paola Gomez Mart\'{\i}nez \footnote{
E-mail address: gomezmar@mdp.edu.ar} , $^{3}$Jos\'e
Edgar Madriz Aguilar\footnote{
E-mail address: jemadriz@fisica.ufpb.br} and
$^{1,2}$Mauricio Bellini\footnote{E-mail address: mbellini@mdp.edu.ar}}

\address{$^1$ Departamento de
F\'{\i}sica, Facultad de Ciencias Exactas y Naturales, Universidad
Nacional de Mar del Plata, Funes 3350, (7600) Mar del Plata,
Argentina.\\
$^2$ Consejo Nacional de Investigaciones
Cient\'{\i}ficas y T\'ecnicas (CONICET). \\
$^3$ Departamento de F\'{\i}sica, Universidade
Federal da Para\'{\i}ba,\\
Caixa Postal 5008, 58059-970 Jo\~{a}o Pessoa, Pb, Brazil.}

\begin{abstract}
We study gravitational waves generated during the inflationary
epoch in presence
of a decaying cosmological parameter on a 5D geometrical background which is Riemann flat. Two examples are
considered, one with a constant cosmological parameter and the
second with a decreasing one.
\end{abstract}

\maketitle

\section{Introduction}

Gravitational waves has been subject of attention since long time
ago \cite{alpha}. Under the standard model of cosmology plus the
theory of inflation, it is very natural to predict the existence
of the background of gravitational waves \cite{1}. Among the
primordial perturbations generated during inflation, there were
basically normal scalar part and tensor part. The primordial
scalar perturbations provided the seeds of large scale structure
which then had gradually formed today's galaxies, which is being
tested in current observations of cosmic microwave background
(CMB). The tensor perturbations have escaped out of the horizon
during inflation, and thus can be completely conserved to form the
relic of background gravitational waves, which carries the
information of the very early universe. In this sense the tensor
perturbations are very significant for studies of very early
universe. Their amplitude is related to the energy scale of
inflation and they are potentially detectable via observations of
$B$-mode polarization in the cosmic microwave background if the
energy scale of inflation is larger than $\sim 3 \times 10^{15}$
GeV \cite{2,3,4,5}. Such a detection would be very important to
test inflation. The direct detection of GW is one of the most
exciting scientific goals because it would improve our
understanding of laws governing the early universe and provide new
means to observe it. The most sensitive detectors which are
already operating, under construction or being planned, are based
on optical interferometers \cite{delta}. In particular,
gravitational wave perturbations will be measured in future by the
Planck satellite, which is designed to produce high-resolution
temperature and polarization maps of CMB. If no primordial GW are detected by CMB, a direct-detection
experiment to understand the simplest form of inflation must have a sensitivity improved by two to three orders of magnitude over 
current plans \cite{*L}. The basal mechanism of
generation of primordial gravitational waves in cosmology has been
discussed in \cite{..}. There are two main 4D formalisms developed
in the literature; the coordinate based approach of
Bardeen \cite{bar} and the covariant formalism \cite{cov}.

The idea that our universe is a 4D space-time embedded in a higher
dimensional has been a topic of increased interest in several
branches of physics, and in particular, in cosmology. This idea
has generated a new kind of cosmological models that includes
quintessential expansion. In particular, theories on which is
considered only one extra dimension have become quite popular in
the scientific community. Among these theories are counted the
braneworld scenarios \cite{bw}, the Space-Time-Matter (STM)
theory \cite{STM} and all noncompact Kaluza-Klein theories.

In this work we shall study the evolution of gravitational waves
on the early universe, which is governed by a (decaying)
cosmological parameter $\Lambda(t)$, from a 5D vacuum state
defined on a $4+1$ Riemann flat spacetime. A decaying cosmological
parameter can be introduced in a geometrical manner through the 5D
background line element \cite{M1}
\begin{equation}\label{eq1}
dS_{b}^{2}=\psi^{2}\frac{\Lambda (t)}{3}dt^{2}-\psi^{2}e^{2\int
\sqrt{\Lambda/3}\,dt}dr^2 - d\psi^{2},
\end{equation}
where $dr^{2}=\delta _{ij}dx^{i}dx^{j}$,
being $\lbrace x^{i} \rbrace =\lbrace x,y,z\rbrace$
the local cartesian coordinates. In addition $t$ and $\psi$ are the time-like and fifth
space-like local coordinates respectively. Adopting a natural
unit system ($\hbar =c=1$), the fifth coordinate $\psi$ has spatial
units whereas the cosmological parameter $\Lambda (t)$ has units of $(length)^{-2}$. The background metric in (\ref{eq1})
is Riemann-flat, $R^{A}\,_{BCD}=0$, describing perfectly a 5D geometric  vacuum.\\

The usual approach would consist to consider the tensor perturbed
line element obtained from (\ref{eq1})
\begin{equation}\label{eq2}
dS^{2}=\psi^{2}\frac{\Lambda (t)}{3}dt^{2}-\psi^{2}e^{2\int
\sqrt{\Lambda/3}\,dt}(\delta _{ij}+Q_{ij})dx^{i}dx^{j}-d\psi ^{2},
\end{equation}
being $Q_{ij}(t,\vec{r},\psi)$ the transverse traceless tensor
denoting the tensor fluctuations with respect to the metric
background (\ref{eq1}), and thereby it satisfies
$tr(Q_{ij})=Q^{i}\;_{i}=0$ and $Q^{ij}_{;i}=0$. The 3D spatial
components of the metric (\ref{eq2}) can be
written as $g_{ij}=-\psi^{2}{\rm exp}\left[2\int \sqrt{\Lambda/3}\,dt\right](\delta _{ij}+Q_{ij})$, so that its contravariant 
components can be linearly
approximated by $g^{ij}\simeq -\psi ^{-2} {\rm exp}\left[-2\int \sqrt{\Lambda/3}\,dt\right](\delta ^{ij}-Q^{ij})$.\\

Under this approach the dynamics obeyed by the tensor fluctuations
$Q_{ij}$, is obtained using the 5D linearized Einstein´s equations in the 5D vacuum $\delta R_{AB}=0$. However, as it is 
well-known
solely
the space-space components contribute for tensor fluctuations, so  the former expression reduces simply to
\begin{equation}\label{eq3}
\delta R_{ij}=0.
\end{equation}
Hence, after some algebra, one obtain the dynamics of the tensor
fluctuations $Q_{ij}$ is determined by
\begin{equation}\label{eq4}
\ddot{Q}_{ij} + \left[3\sqrt{\frac{\Lambda}{3}}-
\frac{1}{2}\frac{\dot{\Lambda}}{\Lambda}\right]\dot{Q}_{ij}-\frac{\Lambda}{3} \  e^{-2\int \sqrt{\Lambda/3}\,dt} \nabla^2_r
Q_{ij}-\frac{\Lambda}{3}\left[4\psi Q_{ij,\psi}+\psi^{2}Q_{ij,\psi,\psi}\right]=0,
\end{equation}
where $(,)$ denotes the partial derivative while the dot is
denoting ordinary derivation with respect the cosmic time $t$.\\

\section{ The 5D tensor modes}

In this letter we shall consider a different approach, which, give
us the dynamics (\ref{eq4}), but from the the action $I=- \int
d^4x \  d\psi \  ^{(5)} L$, given by the background gravitational
Lagrangian, plus free tensorial fields, $Q_{ij}$ ($A$, $B$ run
from $0$ to $4$ and $i$, $j$ from $1$ to $3$)
\begin{equation}
^{(5)} L = \sqrt{\left|\frac{^{(5)} g}{^{(5)} g_0}\right|}
 \left[ \frac{^{(5)} {\cal R}}{16\pi G} + \frac{M^2_p }{2} g^{AB}
Q^{ij}_{\, ;A} Q_{ij;B}\right],
\end{equation}
being $(;)$ the covariant derivative. Furthermore, $^{(5)} {\cal
R}=0$ is the 5D background Ricci scalar and $^{(5)} g = \psi^8
\Lambda e^{6 \int\sqrt{\Lambda/3} dt}/3$ is the determinant of the
background covariant metric tensor $g_{AB}$. Following the usual
quantization process for $Q_{ij}(t,\vec{r},\psi)$
we demand that the next commutation relation must be satisfied
\begin{equation}\label{com}
\left[Q_{ij}(t,\vec{r},\psi),\frac{\partial L}{\partial
Q_{lm,t}}(t,\vec{r'},\psi ')\right]=i \delta^l_i \ \delta^m_j \
 g^{tt} M^2_p \sqrt{\left|\frac{^{(5)}g}{^{(5)}g_0}\right|}
\left(\frac{\psi_0}{\psi}\right)^3
e^{-\int\left[3\sqrt{\frac{\Lambda}{3}} -
\frac{\dot\Lambda}{2\Lambda}\right] dt} \times
\delta^{(3)}(\vec{r}-\vec{r'})\delta(\psi - \psi ').
\end{equation}

On the other hand, $^{(5)} g_0 \equiv ^{(5)} g[\psi=\psi_0,
\Lambda_0\equiv \Lambda(t=t_0)]$, being $\psi_0$ and $t_0$ some
constants to be specified. In this work we are interested to study
the dynamics of $Q_{ij}$. We express the functions $Q_{ij}$ as a
Fourier expansion of the form
\begin{equation}\label{eq5}
Q^{i}\,_{j}(t,\vec{r},\psi)=\frac{1}{(2\pi)^{3/2}}\int
d^{3}k_{r}\,dk_{\psi}\sum _{\alpha}\,^{(\alpha)} e^{i}\,_{j}\left[a_{k_{r}k_{\psi}}^{(\alpha)}
e^{i\vec{k}_{r}\cdot\vec{r}}\zeta _{k_{r}k_{\psi}}(t,\psi)+a_{k_{r}k_{\psi}}^{(\alpha)\,\,\dagger}
e^{-i\vec{k}_{r}\cdot\vec{r}}\zeta _{k_{r}k_{\psi}}^{*}(t,\psi)\right],
\end{equation}
with $\alpha $ counting the number of polarization degrees
of freedom, the asterisk $(*)$ denoting complex conjugate and the creation $a_{k_{r}k_{\psi}}^{(\alpha)\,\,\dagger}$ and 
annihilation
$a_{k_{r}k_{\psi}}^{(\alpha)}$ operators satisfying the algebra
\begin{equation}\label{eq6}
\left[a_{k_{r}k_{\psi}}^{(\alpha)},a_{k'_{r}k'_{\psi}}^{(\alpha ')\,\,\dagger}\right]=g^{\alpha\alpha '}\delta 
^{(3)}(\vec{k}_{r}-\vec{k}'_{r})\delta (\psi -\psi '),\qquad \left[a_{k_{r}k_{\psi}}^{(\alpha)},a_{k'_{r}k'_{\psi}}^{(\alpha 
')}\right]=\left[a_{k_{r}k_{\psi}}^{(\alpha)\,\,\dagger},a_{k'_{r}k'_{\psi}}^{(\alpha ')\,\,\dagger}\right]=0.
\end{equation}
The polarization tensor $^{(\alpha)}e_{ij}$ obeys
\begin{equation}\label{eq7}
^{(\alpha)}e_{ij}=\,^{(\alpha)}e_{ji},\quad ^{(\alpha)}
e_{ii}=0,\quad k^{i}\,^{(\alpha)}e_{ij}=0,\quad ^{(\alpha)}e_{ij}(-\vec{k}_r)=\,^{(\alpha)}e_{ij}^{*}(\vec{k}_r).
\end{equation}
Inserting (\ref{eq5}) in (\ref{eq4}) we obtain that
the dynamics of the 5D tensor modes
$\zeta _{k_{r} m}(t,\psi)$ is given by
\begin{equation}\label{eq8}
\ddot{\zeta}_{k_{r}k_{\psi}}+\left[3\sqrt{\frac{\Lambda}{3}}-\frac{1}{2}
\frac{\dot{\Lambda}}{\Lambda}\right]
\dot{\zeta}_{k_{r}k_{\psi}}+\left[\frac{\Lambda}{3}k_{r}^{2}\,e^{-2\int
\sqrt{\Lambda/3}\,dt}-\frac{\Lambda}{3}\left(4\psi\frac{\partial}{\partial
\psi} +\psi
^{2}\frac{\partial^2}{\partial\psi^2}\right)\right]\zeta_{k_{r}k_{\psi}}=0.
\end{equation}
Decomposing the tensor modes $\zeta _{k_{r}k_{\psi}}(t,\psi)$ into
Kaluza-Klein modes
\begin{equation}\label{eq9}
\zeta _{k_{r}k_{\psi}}(t,\psi)\sim\xi _{k_r}(t)\Theta _{m}(\psi),
\end{equation}
equation (\ref{eq8}) yields
\begin{eqnarray}
\label{eq10}
\ddot{\xi}_{k_r}+\left(3\sqrt{\frac{\Lambda}{3}}-\frac{1}{2}
\frac{\dot{\Lambda}}{\Lambda}\right)\dot{\xi}_{k_r}+\left[\frac{\Lambda}{3
}\,e^{-2\int \sqrt{\Lambda/3}\,dt}k_{r}^{2}+m^{2}\frac{\Lambda}{3}\right]\xi _{k_r}&=&0,\\
\label{eq11} \psi^{2}\frac{d^{2}\Theta
_m}{d\psi^2}+4\psi\frac{d\Theta _{m}}{d\psi}+m^{2}\Theta _{m}&=&0,
\end{eqnarray}
where the parameter $m^2=\left(k_{\psi} \psi\right)^2$ is related
with the squared of the KK-mass measured by a class of observers
in 5D. Using the transformation $\xi _{k_r}(t)=exp\,[-(1/2)\int
(3\sqrt{\Lambda/3}-(\dot{\Lambda}/2\Lambda))\,dt]\,\chi _{k_r}(t)$
and $\Theta_m(z) = e^{-3/2 z} L_m(z)$, with $z={\rm
ln}(\psi/\psi_0)$, in eqs. (\ref{eq10}) and (\ref{eq11}),
respectively, we obtain
\begin{eqnarray}
&& \ddot{\chi}_{k_r}+\left[\frac{\Lambda}{3}\,e^{-2\int \sqrt{\Lambda/3}\,dt} k_{r}^{2}-
\frac{1}{4}\sqrt{\frac{3}{\Lambda}}\dot{\Lambda}+\frac{1}{4}
\frac{\ddot{\Lambda}}{ \Lambda}-\frac{5}{16}\frac{\dot{\Lambda}^2}{\Lambda^2}
+\frac{3}{4}\sqrt{\frac{\Lambda}{3}}\frac{\dot{\Lambda}}{\Lambda}+\left( \frac{m^2}{3}-\frac{3}{4}\right)\Lambda \right]\chi 
_{k_r}=0, \label{eq12}\\
&& \frac{d^2 L_m(z)}{dz^2} + \left[m^2 - \frac{9}{4}\right] L_m(z) =0.
\label{eq12'}
\end{eqnarray}
This way, given a cosmological parameter $\Lambda (t)$, the
temporal evolution of the tensor modes $\xi _{k_r}(t)$ in 5D is determined by solutions of (\ref{eq12}). Once a
solution for $\xi _{k_r}(t)$ is obtained, it should satisfy the
algebra (\ref{com}). This can be made guaranteeing that such a solution obey
\begin{equation}\label{eq13}
\chi_{k_{r}}\dot{\chi}_{k_{r}}^{*}-
\dot{\chi}_{k_{r}}\chi_{k_{r}}^{*}=i, \qquad \left| L_m\right|^2=1.
\end{equation}
On the other hand, note that equation (\ref{eq11}) is exactly the
same as the one obtained in \cite{Ed}.
Therefore about the behavior of the modes with respect the
fifth coordinate we can say that for $m>3/2$ the KK-modes are coherent on the ultraviolet sector (UV), described by the modes
\begin{equation}\label{eq14}
k^2_r > \left\{\frac{3}{2\Lambda} \frac{d}{dt}\left[ 3
\sqrt{\frac{\Lambda}{3}} - \frac{\dot\Lambda}{2\Lambda}\right]
-\frac{3}{4\Lambda}
\left(3\sqrt{\frac{\Lambda}{3}}-\frac{\dot\Lambda}{2\Lambda}\right)^2
- m^2\right\} \  e^{2\int \sqrt{\frac{\Lambda}{3}} dt} >0.
\end{equation}
However, for $m<3/2$ those modes are unstable and diverge at
infinity. The modes with $m=0$ and $m > 3/2$ comply with the
conditions
(\ref{eq13}), so that they are normalizable.\\

\section{Effective 4D dynamics}

To describe the 4D dynamics we can make a foliation on
$\psi=\psi_0$ on the line element (\ref{eq1}), such that the
effective 4D background metric holds: $\left.dS^2\right|_{eff} =
ds^2$, where
\begin{equation}\label{61}
ds^2 =
\psi^2_0 \frac{\Lambda(t)}{3} dt^2 - \psi^2_0 e^{2 \int
\sqrt{{\Lambda\over 3}}dt} dr^2.
\end{equation}
In this section we shall study the dynamics of the 4D
tensor-fluctuations $h_{ij}(t,\vec{r})\equiv
Q_{ij}(t,\vec{r},\psi=\psi_0)$, making emphasis on the long
wavelength section, which describes this field on cosmological
scales. The effective 4D action $^{(4)} I$ is ($\alpha$, $\beta$
run from $0$ to $3$)
\begin{equation}\label{act}
^{(4)} I = -\int d^4 x \sqrt{\left|\frac{^{(4)} g}{^{(4)}
g_0}\right|} \left. \left[ \frac{^{(4)} {\cal R}}{16\pi G} +
\frac{M^2_p }{2} g^{\alpha\beta}  Q^{ij}_{\, ;\alpha}
Q_{ij;\beta}\right] \right|_{\psi=\psi_0},
\end{equation}
where $^{(4)} {\cal R} = 12/\psi^2_0$ is the effective 4D Ricci
scalar valuated on the metric (\ref{61}), such that we obtain an
equation of state which describes an effective 4D vacuum: ${\rm p}
= -\rho$, being ${\rm p}$ and $\rho$ the pressure and the energy
density, respectively. Hence, the metric (\ref{eq1}) could be
considered as an extension of the Ponce de Leon one\cite{PdL},
which, on a foliation $\psi=\psi_0$, also describes an effective
4D vacuum dominated expansion.

The effective 4D equation of motion for the 4D
tensor-fluctuations is
\begin{equation}\label{eq17}
\ddot{h}^i_j + \left[3\sqrt{\frac{\Lambda}{3}} -
\frac{\dot\Lambda}{2\Lambda} \right] \dot{h}^i_j -
\frac{\Lambda}{3} e^{-2\int\sqrt{\frac{\Lambda}{3}} dt} \nabla^2_r
h^i_j - \left. \frac{\Lambda}{3} \left[\frac{4}{\psi}
\frac{\partial Q^i_j}{\partial \psi} + \psi^2 \frac{\partial^2
Q^i_j}{\partial\psi^2} \right]\right|_{\psi= \psi_0} =0.
\end{equation}
Using the eq. (\ref{eq11}), we obtain
\begin{equation}\label{eq18}
\ddot{h}^i_j + \left[3\sqrt{\frac{\Lambda}{3}} -
\frac{\dot\Lambda}{2\Lambda} \right] \dot{h}^i_j -
\frac{\Lambda}{3} e^{-2\int\sqrt{\frac{\Lambda}{3}} dt} \nabla^2_r
h^i_j + \frac{m^2 \Lambda}{3} h^i_j =0.
\end{equation}
After make the transformation $h^i_j(t,\vec r) = e^{-1/2 \int \left[
3\left({\Lambda\over 3}\right)^{1/2}-{\dot\Lambda\over 2\Lambda}\right] dt}
\chi^i_j(t,\vec r)$, we obtain
\begin{equation}\label{eq19}
\ddot{\chi}^i_j - \frac{\Lambda}{3}
e^{-2\int\sqrt{\frac{\Lambda}{3}} dt} \nabla^2_r \chi^i_j - \left[
\frac{1}{4} \sqrt{\frac{3}{\Lambda}} \dot\Lambda + \frac{1}{4}
\frac{\ddot\Lambda}{\Lambda} - \frac{5}{16}
\frac{\dot\Lambda^2}{\Lambda^2} + \frac{3}{4}
\sqrt{\frac{\Lambda}{3}} \frac{\dot\Lambda}{\Lambda} + \left(
\frac{m^2}{3} - \frac{3}{4} \right)  \Lambda \right] \chi^i_j =0,
\end{equation}
such that it is possible to make a Fourier expansion for $\chi^i_j$
\begin{equation}\label{eq5'}
\chi^{i}\,_{j}(t,\vec{r})=\frac{1}{(2\pi)^{3/2}}\int
d^{3}k_{r}\,dk_{\psi}\sum _{\alpha=+,\times}\,^{(\alpha)} e^{i}\,_{j}\left[a_{k_{r}k_{\psi}}^{(\alpha)}
e^{i\vec{k}_{r}\cdot\vec{r}}\chi _{k_{r}k_{\psi}}(t,\psi)+a_{k_{r}k_{\psi}}^{(\alpha)\,\,\dagger}
e^{-i\vec{k}_{r}\cdot\vec{r}}\chi _{k_{r}k_{\psi}}^{*}(t,\psi)\right] \delta(k_{\psi}-k_{\psi_0}),
\end{equation}
where we require that the modes $\chi_{k_r}\equiv \chi_{k_r k_{\psi_0}}$ satisfy the commutation relation
\begin{equation}\label{cr1}
\left[\chi_{k_{r}}(t,\vec{r}),\dot{\chi}_{k_r}(t,\vec{r'})\right]=i\delta^{(3)}(\vec{r}-\vec{r'}).
\end{equation}
Using (\ref{eq5'}) this expression reads
\begin{equation}\label{mod}
\chi_{k_{r}} \dot\chi^*_{k_{r}} - \chi^*_{k_{r}} \dot\chi_{k_{r}}=i,
\end{equation}
which is the condition for the modes to be normalizable on the UV-sector. Inserting (\ref{eq5'}) in (\ref{eq19}) we obtain the 
dynamical equation for the $k_{r}$-modes
\begin{equation}\label{deq1}
\ddot{\chi}_{k_{r}}+\left[\frac{\Lambda}{3}e^{-2\int\sqrt{\frac{\Lambda}{3}}dt}k_{r}^{2}-\left(\frac{1}{4}\sqrt{\frac{3}{\Lambda}}
\dot{\Lambda}+\frac{1}{4}\frac{\ddot{\Lambda}}{\Lambda}-\frac{5}{16}\frac{\dot{\Lambda}^{2}}{\Lambda^2}+\frac{3}{4}\sqrt{ \frac{ 
\Lambda}{3}}\frac{\dot{\Lambda}}{\Lambda}+\left(\frac{m^{2}}{3}-\frac{3}{4}\right)\Lambda\right)\right]\chi _{k_r}=0.
\end{equation}
This way for a given $\Lambda (t)$ corresponds a normalized solution for the $k_{r}$-modes by solving (\ref{deq1}).  Once obtained 
a normalized solution of (\ref{deq1}), we will be able of dealing with the spectrum on super Hubble scales. The amplitude of the 
4D tensor metric fluctuations $<h^{2}>=<0|h^{i}\,_{j}\,h_{i}\,^{j}|0>$ on the IR-sector is given by
\begin{equation}\label{deq2}
\left<h^{2}\right> =\frac{4}{\pi^2}e^{-\int \left[
3\left({\Lambda\over 3}\right)^{1/2}-{\dot\Lambda\over 2\Lambda}\right] dt}\int _{0}^{\epsilon 
k_H}\frac{dk_{r}}{k_r}k_{r}^{3}\left[\chi _{k_r}(t)\chi _{k_r}^{*}(t)\right]_{IR},
\end{equation}
where $\epsilon=k_{max}^{IR}/k_p \ll 1$ is a dimensionless parameter, being $k_{max}^{IR}=k_{H}(t_i)$ the wave-number related to 
the Hubble radius at the time $t_{i}$. This time corresponds at the time when the gravitational modes re-enter to the horizon. In 
addition, $k_p$ is the Planckian wave-number. Clearly, in order to obtain an explicit spectrum it is necessary to specify first a 
functional form for $\Lambda (t)$. Some illustrative examples will be studied in the next section.

\section{Examples}

In order to illustrate the formalism developed in the previous section, along the  present section we study a pair of interesting 
examples. The first one contemplating a constant cosmological parameter
$\Lambda=3H_{0}^{2}$, and the second one considering a decaying  $\Lambda (t)=3p^{2}/t^{2}$.

\subsection{Case $\Lambda=3 H^2_0$}

The simplest example results of taking the cosmological parameter $\Lambda$ to be a
constant, and in particular $\Lambda = 3 H^2_0$. In this particular case the
equation of motion for the modes $\chi_{k_{r}}$ becomes
\begin{equation}\label{modd}
\ddot\chi_{k_r} + \left[H^2_0 e^{-2 H_0 t} k^2_r-
\left(m^2 - \frac{9}{4}\right) H^2_0 \right] \chi_{k_r} =0.
\end{equation}
The general solution for this equation is
\begin{equation}\label{sol}
\chi_{k_r}(t) = A_1 \  {\cal H}^{(1)}_{\nu}\left[k_r e^{-H_0t}\right]
+ A_2 \  {\cal H}^{(2)}_{\nu}\left[k_r e^{-H_0 t}\right],
\end{equation}
where $A_{1}$ and $A_{2}$ are integration constants. After the Bunch-Davies normalization \cite{BD} the normalized solution reads
\begin{equation}\label{so1}
\chi_{k_r}(t) = \frac{i}{2}\sqrt{\frac{\pi}{H_{0}}}
{\cal H}^{(2)}_{\nu}\left[k_r e^{-H_0 t}\right],
\end{equation}
with $\nu = (1/2)\sqrt{4m^2 - 9}$. This solution is stable
for $m^2 > 9/4$,
on scales $k^2_r > \left(m^2 - {9\over 4}\right) e^{2 H_0 t} >0$. Now considering the asymptotic expansion for the Hankel function 
${\cal H}_{\nu}^{(2)}[y]\simeq (-i/\pi)\Gamma (\nu)[y/2]^{-\nu}$ in (\ref{so1}), the amplitude of the 4D tensor metric 
fluctuations (\ref{deq2}) on cosmological (super Hubble) scales
gives
\begin{equation}\label{deq3}
\left<h^{2}\right>_{SH}
=\frac{2^{2\nu}}{\pi^{3}(3-2\nu)}\frac{\Gamma^{2}(\nu)}{H_{0}}e^{-(3-2\nu)H_{0}t}\epsilon ^{3-2\nu}k_{H}^{3-2\nu},
\end{equation}
where $k_{H}(t)\sim \, e^{H_{0}t}$. Hence, the gravitational
spectrum ${\cal P}_{g}(k_r)$ takes the form
\begin{equation}\label{deq4}
{\cal P}_{g}(k_r)=
\left.\frac{2^{2\nu}}{\pi^{3}}\frac{\Gamma^{2}(\nu)}{H_0}e^{-(3-2\nu)H_{0}t}k_{r}^{3-2\nu}\right|_{k_r=\epsilon
k_{H}}
\end{equation}
We can see from (\ref{deq4}) that for $m\simeq 3/\sqrt{2}>(3/2)$,
the gravitational spectrum ${\cal P}_{\nu}(k_r)$ is nearly scale invariant and consequently the tensor spectral index becomes 
$n_{T}\equiv 3-2\nu\simeq 0$ in this case.

\subsection{Case $\Lambda = 3 p^2/t^2$}

Another interesting case appears considering a decaying cosmological
parameter $\Lambda = 3 p^2/t^2$, with the restriction $\dot\Lambda <0$. In
this case the equation of motion for the modes $\chi_{k_r}(t)$
results to be
\begin{equation}\label{ecuac}
\ddot\chi_{k_r} + \left\{ k^2_r p^2 t^{2p}_0 t^{-2(p+1)} -
\left[\left(m^2-\frac{9}{4}\right)p^2 -\frac{9}{4} p +
\frac{1}{4}\right] t^{-2} \right\} \chi_{k_r} =0,
\end{equation}
where $M^2(t) = \left[\left(m^2 - {9\over 4}\right) p^2 - {9\over
4} p + {1\over 4} \right] t^{-2}$ can be interpreted as an
effective squared term of mass. The permitted values for $m$
should be
\begin{equation}\label{m}
\frac{9}{4} < m^2 \leq \frac{9}{4} \left(\frac{9}{4} +1\right),
\end{equation}
for which
\begin{equation}\label{p}
\frac{9 - \sqrt{117 - 16 m^2}}{2(4 m^2-9)} \leq p \leq \frac{9 +
\sqrt{117 - 16 m^2}}{2(4 m^2-9)}.
\end{equation}
The general solution of (\ref{ecuac}) can be expressed in terms of Bessel functions as
\begin{equation}\label{ecu}
\chi_{k_r}(t) = B_1 \left[\frac{y(t)}{2}\right]^{-\mu/(2p)}t^{(1-\mu)/2}\Gamma\left(1+\frac{\mu}{2p}\right){\cal 
J}_{\mu}[y(t)]+B_{2}\left[\frac{y(t)}{2}\right]^{\mu/(2p)}t^{(1+\mu)/2}\Gamma\left(1-\frac{\mu}{2p}\right){\cal J}_{-\mu}[y(t)],
\end{equation}
where $\mu = \sqrt{4p^{2}m^{2}-9p(p+1)+2}$ and $y(t)=k_{r}(t_{0}/t)^{p}$. In general this expression is not
normalizable. However there exist some particular solutions of (\ref{p}) that are normalizable. A particular case that yields a 
normalizable solution results by taking $m=\pm [1/(2p)]\sqrt{9p(p+1)-1}$. In this case
the dynamical equation for the quantum gravitational modes reduces
to
\begin{equation}\label{deq5}
\ddot{\chi} _{k_r}+k_{r}^{2}p^{2}t_{0}^{2p}t^{-2(p+1)}\chi _{k_r}=0,
\end{equation}
whose general solution is
\begin{equation}\label{deq6}
\chi _{k_r}(t)=\sqrt{t}\left\{C1{\cal H}_{\omega}^{(1)}\left[k_{r}\left(\frac{t_0}{t}\right)^{p}\right]+C_{2}{\cal 
H}_{\omega}^{(2)}\left[k_{r}\left(\frac{t_0}{t}\right)^{p}\right]\right\},
\end{equation}
being $C_{1}$ and $C_{2}$ integration constants, and $\omega
=1/(2p)$. The Bunch-Davies normalized solution is then
\begin{equation}\label{deq7}
\chi _{k_r}(t)=\frac{i}{2}\sqrt{\frac{1}{\pi p}}\,{\cal H}_{\omega}^{(2)}\left[k_{r}\left(\frac{t_0}{t}\right)^{p}\right].
\end{equation}
In this case the amplitude of the 4D tensor metric fluctuations on
super-Hubble scales ($k_r \gg k_H$) reads
\begin{equation}\label{deq8}
\left<h^{2}\right>_{SH} =\frac{2^{2\omega}}{p\,\pi 
^{5}}\frac{\Gamma^{2}(\omega)}{3-2\omega}\left(\frac{t_0}{t}\right)^{(3-2\omega)p+1}\epsilon^{3-2\omega}k_{H}^{3-2\omega}.
\end{equation}
Therefore the gravitational spectrum ${\cal P}_{g}(k_{r})$ is in this case
\begin{equation}\label{deq9}
{\cal P}_{g}(k_r)=\left.\frac{2^{2\omega}}{p\,\pi^{5}}\Gamma^{2}(\omega)\left(\frac{t_0}{t}\right)^{(3-2\omega)p+1}
k_{r}^{3-2\omega} \right|_{k_r=\epsilon k_H}\,.
\end{equation}
Clearly, for $p\simeq 1/3$ (that corresponds to $m \simeq
3\sqrt{3}/2 > 3/2$), the spectrum is nearly scale invariant i.e.
$n_{T}\equiv 3-2\omega \simeq 0$.

\section{Final Comments}

In this letter we have studied the emergence of gravitational
waves in the early universe, which is considered as dominated by a
decaying cosmological parameter, from a 5D Riemann flat background
metric, on which we define the 5D vacuum. Our approach is
different to others, because we consider gravitational waves as
originated by a physical field $Q_{ij}$, but not a tensorial
linearized fluctuation of the metric. Therefore, it is possible to
deal with $Q_{ij}$ as a quantum traceless tensorial field with
null divergence.

The effective 4D dynamics of GW, $h_{ij} = Q_{ij}(\psi = \psi_0)$,
can be viewed as induced by the foliation of the fifth coordinate.
In this letter we have worked two examples: {\bf a)} In the case
with constant cosmological parameter, $\Lambda = \Lambda_0$, we
obtain that the KK mass of the gravitons should be $m \simeq
3/\sqrt{2}$ to obtain a nearly scale invariant tensorial power
spectrum of $\left<h^2\right>_{SH}$. This value of $m$ corresponds
to an effective 4D mass $M \simeq 3H_0/2 = \sqrt{3\Lambda}/2$.
Note that this mass can be related to the Einstein mass $m_E$, as
it is considered in\cite{wess}: $M \simeq {3\over 2 G m_E}$. {\bf
b)} The case where the cosmological parameter decrease with time
is the more interesting. In this case the gravitons are
normalizable only when their effective 4D mass $M$ become null. We
obtain that the spectrum of the squared fluctuations of
gravitational waves are scale invariant for KK masses with values
$m\simeq 3\sqrt{3}/2$.

\vskip .2cm
\centerline{\bf{Acknowledgements}}
\vskip .2cm
SPGM acknowledges UNMdP for financial support.
MB acknowledges CONICET
and UNMdP (Argentina) for financial support.
JEMA acknowledges CNPq-CLAF and UFPB for financial support.
\\

\end{document}